\documentclass[aps,prd,showpacs,preprintnumbers,twocolumn,groupedaddress]{revtex4}
\usepackage[dvipdfmx]{color}
\usepackage[dvipdfmx]{graphicx}
\usepackage{amsmath}
\usepackage{color}

\usepackage[normalem]{ulem}  
\renewcommand\sout{\bgroup \color{red} \ULdepth=-.5ex \ULset}

\begin{document}

\title{ Nambu-Goldstone bosons and the Higgs mechanism without Lorentz invariance: Analysis based on constrained-system theory }

\author{Shinya~Gongyo$^{1,2}$ and Shintaro~Karasawa$^{1}$}

\affiliation{
$^1$Department of Physics, Kyoto University, Kyoto 606-8502, Japan\\
$^2$Department of Physics, New York University, New York, 10003, USA}

\date{\today}

\begin{abstract}
First, we develop the counting rule for Nambu-Goldstone bosons to the system including two-time derivative terms. In this case, the type-II Nambu-Goldstone bosons may appear along with the massive Nambu-Goldstone ones. The number of the bosons is not reduced in contrast to the system without two-time derivative terms. We also investigate the reduction of the degrees of freedom from the perspective of the Dirac-Bergmann theory of constraints and reproduce the counting rule for Nambu-Goldstone bosons without Lorentz invariance. Then, we construct the generic Higgs model and study a Higgs phenomenon with these Nambu-Goldstone bosons on the basis of the Dirac-Bergmann theory. We show that the gauge fields in this system absorb all of the Nambu-Goldstone bosons such as the type-I, type-II and massive ones.
\end{abstract}


\maketitle

\section{Introduction}
 Spontaneous symmetry breaking and the Higgs mechanism are well-known phenomena in a wide area ranging from the particle physics to condensed matter physics \cite{Nambu:1961tp,Goldstone:1961eq,Higgs:1964ia}. Through Spontaneous symmetry breaking, Nambu-Goldstone bosons appear in both Lorentz invariant and noninvariant systems. In the Lorentz invariant case, the dispersion of Nambu-Goldstone bosons is always linear and the number of Nambu-Goldstone bosons coincides with that of the generators of the broken symmetries, while in the noninvariant case, the dispersion of Nambu-Goldstone bosons may be nonlinear and the number of Nambu-Goldstone bosons does not alway coincide with that of the broken symmetries. \cite{Nielsen:1975hm,Hidaka:2012ym,Watanabe:2012hr,Leutwyler:1993gf,Schafer:2001bq} 

Nielsen and Chadha performed the classification of these Nambu-Goldstone bosons into type-I and type-II Nambu-Goldstone bosons whose dispersions are characterized by odd and even powers of the momentum, respectively and showed an inequality for the relation between these bosons \cite{Nielsen:1975hm}. Recently a general counting rule for these bosons has been given based on Mori's projection operator method by Hidaka \cite{Hidaka:2012ym} and Leutwyler's effective Lagrangian by Watanabe and Murayama \cite{Watanabe:2012hr, Leutwyler:1993gf}. Furthermore, in addition to these Nambu-Goldstone bosons, the appearance of massive Nambu-Goldstone bosons whose dispersion is finite at zero momentum is reported in Refs. \cite{Hidaka:2012ym,Kapustin:2012cr,Nicolis:2012vf,Watanabe:2013uya,Gusynin:2003yu,Hama:2011rt}. The relation between the presence of two-time derivative terms in the effective Lagrangian and massive Nambu-Goldstone bosons was pointed out by Kapustin \cite{Kapustin:2012cr}.

Meanwhile, to our knowledge, the Higgs phenomenon with these Nambu-Goldstone bosons has not been discussed generally except for some model analysis \cite{Gusynin:2003yu,Hama:2011rt}. 
From the theoretical viewpoint, the number of ``eaten" degrees of freedom by the gauge fields is of interest. Whether the degrees of freedom correspond to the number of Nambu-Goldstone bosons or broken symmetries in the non-Lorentz invariant systems is not clear in contrast to the Lorentz invariant systems. 

One of the situations of the Higgs phenomenon without Lorentz invariance is the SU(2) Higgs-Kibble model with finite chemical potential. In this model, the Higgs phenomenon with type-I, type-II and massive Nambu-Goldstone bosons occurs \cite{Gusynin:2003yu,Hama:2011rt}. In addition, another nonrelativistic Higgs-phenomenon with these Nambu-Goldstone bosons may also appear in the context of ultracold atoms \cite{Dalibard:2010ph}.

In Sec. \ref{counting}, we start from the effective Lagrangian approach keeping two-time derivative terms and develop a general counting rule for type-I, type-II and massive Nambu-Goldstone bosons. In Sec.\ref{constraints for NG} we identify the physical degrees of freedom for this system including two-time derivative terms based on the Dirac-Bergmann theory for constraints. Then, in Sec.\ref{constraints for Higgs}, we study the Higgs phenomenon without Lorentz invariance and clarify the number of absorbed physical degrees of freedom by the gauge fields. Finally, we show some examples about the Higgs model without Lorentz invariance in Sec.\ref{examples}.

\section{Counting rule for type-I, type-II and massive Nambu-Goldstone bosons}
\label{counting}
First, we extend a counting rule for type-I and type-II Nambu-Goldstone bosons in Ref. \cite{Watanabe:2012hr} to the case including the term with two-time derivatives in the effective Lagrangian without Lorentz invariance \cite{Kapustin:2012cr}. We also discuss the influence of two-time derivative terms on the dispersion of these Nambu-Goldstone bosons.
The non-relativistic effective Lagrangian based on Spontaneous symmetry breaking $G \rightarrow  H $ is given by
\begin{align}
\mathcal{L}_{\pi} &= c_a(\pi) \dot{\pi}^a + \frac{1}{2} \bar{g}_{ab} (\pi) \dot{\pi}^a \dot{\pi}^b - \frac{1}{2} g_{ab} (\pi) \partial_{r} \pi^a \partial_{r} \pi^b \nonumber \\
&= \partial_a c_b |_{\pi=0} \pi ^a \dot{\pi}^b + \frac{1}{2} \bar{g}_{ab} (0) \dot{\pi}^a \dot{\pi}^b - \frac{1}{2} g_{ab} (0) \partial_{r} \pi^a \partial_{r} \pi^b \nonumber \\
&+O(\pi ^3)
 \label{k2}
\end{align}
with $\pi ^a~(a=1, ... ,\mathrm{dim} G/H)$ pion fields, $\partial_a \equiv \partial / \partial \pi ^a$ the derivative with respect to the pion fields and $r$ the spatial directions \cite{Leutwyler:1993gf}. Here we do not assume that $\bar{g}_{ab}(0)$ is regular in contrast to Ref. \cite{Kapustin:2012cr}.

The infinitesimal transformation of pion fields is given by $\delta  \pi ^a = \epsilon ^i h_i ^a (\pi)~(i = 1, ...,\mathrm{dim} G)$ and the relations between the coefficients of derivative terms, $c_a (\pi), \bar{g}_{ab} (\pi), g_{ab}(\pi)$, and $h_i ^a (\pi)$ are obtained in Refs. \cite{Leutwyler:1993iq, Leutwyler:1993gf}. Note that the pion fields can be chosen to transform linearly under the unbroken symmetry $H$ and thus $h^a_{i} (0)=0$ for the indices of  the unbroken symmetry. $g_{ab}(\pi)$ corresponds to a metric and one of the coefficients $h_{ai} (\pi)$ in a gauged system discussed later is expressed by $h_{ai} (\pi) = g_{ab} \left(\pi\right)h^b_i \left(\pi \right)$. Following Leutwyler's notation, we use $i,j,k...=1,... \mathrm{dim G}$ and $a,b,c,...=1,... \mathrm{dim}G/H$ \cite{Leutwyler:1993iq, Leutwyler:1993gf}.


The first term in Eq.(\ref{k2}) is related to the commutation relations between conserved charges, 
\begin{align}
\rho_{ij} \equiv -i \lim_{\Omega \to \infty} \frac{1}{\Omega} \langle 0 | [{Q}_i, {Q}_j] | 0 \rangle
\end{align}
with $Q_i $ the Noether charge and $\Omega$ the spatial volume of the system \cite{Watanabe:2012hr},
\begin{align}
h^a_i h^b_j\left( \partial _b c_a - \partial_a c_b \right)|_{\pi =0}= \rho_{ij}.
\end{align}
By neglecting the total derivative terms and changing the pion fields into $\tilde{\pi}^a \equiv \pi ^b \left(h ^{-1}\right)_b^{a}$, the Lagrangian is reduced to 
\begin{align}
\mathcal{L}_{\pi} = \frac{1}{2} \rho_{ab} \dot{\tilde{\pi}}^a \tilde{\pi}^b -\frac{1}{2} \tilde{\bar{g}}_{ab} (0) \tilde{\pi}^a \ddot{\tilde{\pi}}^b + \frac{1}{2} \tilde{g}_{ab} (0) \tilde{\pi}^a \partial_r^2 \tilde{\pi}^b + \mathcal{O}(\tilde{\pi}^3), \label{k4}
\end{align}
where $\tilde{\bar{g}}_{ab}(0)$ and $\tilde{g}_{ab} (0)$ are defined as
\begin{align}
&\tilde{\bar{g}}_{ab} (0) = \bar{g}(0)_{cd} h(0)_{a}^c h(0)_{b}^d, \nonumber \\
&\tilde{{g}}_{ab} (0)= g(0)_{cd} h(0)_{a}^c h(0)_{b}^d. \label{g_gbar}
\end{align}

In the relativistic case, $\tilde{\bar{g}}_{ab} (0)=\tilde{{g}}_{ab} (0)$ is diagonal and connected to the pion decay constants, $\tilde{{g}}_{ab} (0)=\delta _{ab} F^2_a$ \cite{Leutwyler:1993iq}. In the nonrelativistic case, two kinds of the pion decay constants, $F_a$ and $\bar{F}_a$ appear, corresponding to the time and spatial components. By analogy with the relativistic case, $\tilde{\bar{g}}_{ab} (0) $ and $\tilde{{g}}_{ab} (0)$ are expected to be related to these pion decay constants as follows:
\begin{align}
&\tilde{\bar{g}}_{ab} (0) = \delta_{ab}\bar{F}_a^2, \nonumber \\
&\tilde{{g}}_{ab} (0)= \delta _{ab} F^2_a, \label{pion_decay_const}
\end{align}
where the summation for $a$ is not taken. Whether pion decay constants are degenerate or not depends on the structure of $G$ and $H$. The number of undegenerate pion decay constants corresponds to that of representations of $G/H$ transforming under $H$ irreducibly.

Note that we assume $\mathrm{rank} g(0) =\mathrm{dim} G/H$, because if ${\rm rank}g (0) < {\rm dim} G/H$, some pion fields without quadratic spatial derivative terms appear in the terms only up to $O(\pi ^2)$ .
 On the other hand, $\mathrm{rank} \bar{g}(0)  \leq G/H$ may occur, corresponding to the case with $\bar{F}_a^2=0$ for some $a$. 

Next we perform the transformation which reduces $\rho_{ab}$ to the block diagonal form by an orthogonal matrix. The block diagonal form is composed of three sectors classified by time derivatives. The one characterized by $\hat{u}_\alpha \left(\alpha = 1, \dots ,p\right)$ includes one- and two-time derivatives, the one characterized by $\hat{v}_\beta \left(\beta = 1, \dots ,q\right)$ just one-time derivative and the one characterized by $w_\gamma \left(\gamma = 1, \dots ,s\right)$ just two-time derivatives as follows:

\begin{align}
&\mathcal{L}_{\pi} = \frac{1}{2} \tilde{\pi}'^{\dagger} A \tilde{\pi}', \label{k6} \\
&A= \left(
\begin{array}{ccccccccc}
	\hat{u}_1 &&&&&&&& \\
	&\ddots&&&&&&& \\
	&&\hat{u}_p&&&&&& \\
	&&&\hat{v}_{1}&&&&& \\
	&&&&\ddots&&&& \\
	&&&&&\hat{v}_{q}&&& \\
	&&&&&&w_{1}&& \\
	&&&&&&&\ddots& \\
	&&&&&&&&w_s \\
\end{array}
\right), \label{k7} \\
&\hat{u}_\alpha = \left(
\begin{array}{cc}
	- \bar{F}^2_{\alpha_{(1)}}\partial _0 ^2 + F^2_{\alpha_{(1)}}\partial _r ^2 & -\theta_\alpha \partial_0 \\
	\theta_\alpha \partial_0 & -\bar{F}^2_{\alpha_{(2)}}\partial _0 ^2 + F^2_{\alpha_{(2)}}\partial _r ^2 \\
\end{array}
\right)
, \nonumber \\
&\hat{v}_\beta = \left(
\begin{array}{cc}
	F^2_{\beta _{(1)}}\partial_r^2 & -\theta_{p+\beta} \partial_0 \\
	\theta_{p+\beta} \partial_0 & F^2_{\beta_{(2)}}\partial_r^2 \\
\end{array}
\right) , \nonumber \\
&w_\gamma= -\bar{F}^2_{2p+2q+\gamma}\partial _0 ^2 + F^2_{2p+2q+\gamma}\partial _r ^2,
 \label{k8}
\end{align}
where we assume that $\bar{F}_{2p+1}^2=...=\bar{F}_{2p+2q}^2=0$ in order to consider the case of $\mathrm{rank} \tilde{\bar{g}}(0)<G/H$. $\theta_\alpha$ and $\theta_{p+\beta}$ are real components for the block diagonal form of $\rho_{ab}$. 
Notations in Eqs.(\ref{k7}) and (\ref{k8}) are given as follows: $\alpha _{(1)}=2\alpha -1,\alpha _{(2)}=2\alpha, \beta _{(1)}=2p+2\beta -1$ and $\beta _{(2)}=2p+2\beta$ denote the indices in each $\hat{u}_\alpha$ and $\hat{v}_\beta$ sector. $2p={\rm rank}\tilde{\bar{g}}(0) + {\rm rank} \rho - {\rm dim} G/H$, $2q={\rm dim} G/H - {\rm rank} \tilde{\bar{g}}(0)$ and $s=\mathrm{dim}G/H - \mathrm{rank} \rho$ and thus $2p+2q+s=\mathrm{dim}G/H$ is satisfied.

Due to this block diagonal form of the matrix $A$, each of these sectors leads to the dispersion relations:
\begin{align}
E_\alpha^u =& \Bigg \{
\begin{array}{lll}
	\frac{F_{\alpha_{(1)}}F_{\alpha_{(2)}}}{\theta_\alpha} p^2 + \mathcal{O}(p^4)&& \\
	&& \\
	\frac{1}{\bar{F}_{\alpha_{(1)}} \bar{F}_{\alpha_{(2)}}}\theta_\alpha+ \frac{F_{\alpha_{(1)}}^2 \bar{F}_{\alpha_{(2)}}^2 + \bar{F}_{\alpha_{(1)}}^2 F_{\alpha_{(2)}}^2}{2 \theta_\alpha \bar{F}_{\alpha_{(1)}} \bar{F}_{\alpha_{(2)}}}p^2  + \mathcal{O}(p^4),&& \\
\end{array} \nonumber \\
E_\beta^v =& \frac{F_{\beta_{(1)}}F_{\beta_{(2)}}}{\theta_\beta} p^2, && \nonumber \\
E_{\gamma}^w=& \frac{F_{2p+q+\gamma}}{\bar{F}_{2p+q+\gamma}}p. \label{k10}
\end{align}
We find that type-II and massive Nambu-Goldstone modes appear in the sector of $\hat{u}_\alpha$, while only type-II Nambu-Goldstone modes appear in the sector of $\hat{v}_\beta$ corresponding to the one reported in Ref. \cite{Watanabe:2012hr}.
The $w_\gamma$ sector gives type-I Nambu-Goldstone modes.

In $\hat{v}_\beta$ sector, each pion field connected with one time derivative is regarded as canonical variables. Thus the physical degrees of freedom in this sector are halved and $q$ type-II Nambu-Goldstone modes are realized as clarified later with the Dirac-Bergmann theory.
On the other hand, in $\hat{u}_\alpha$ sector, the type-II modes are accompanied by massive modes and the degrees of freedom are not reduced in general.

Now let us discuss the relation between the $\hat{u}_\alpha$ and $\hat{v}_\beta$ sectors. Here, for simplicity we assume that each of time- and spatial- pion decay constants do not depend on the group index in the sector of $\hat{u}_\alpha$, $\bar{F}$ and $F$. By using Eqs.(\ref{k4}) and (\ref{pion_decay_const}), the two-time derivative terms are neglected in the region of $E\ll \left(F/\bar{F}\right)p$. In this region, type-II modes are always found, while massive Nambu-Goldstone modes are not found and the physical degrees of freedom are halved effectively, which corresponds to the case analyzed by Hidaka and Watanabe and Murayama \cite{Hidaka:2012ym,Watanabe:2012hr}.

\begin{figure*}[h]
\begin{center}
\includegraphics[scale=0.4]{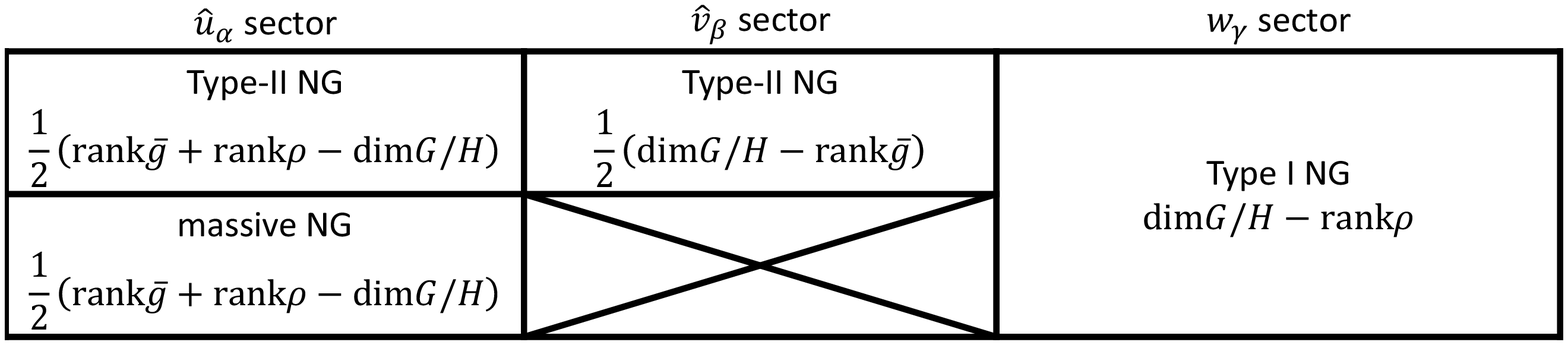}
\caption{\label{Fig1}
The counting rule for type-I, type-II and masive Nambu-Goldstone bosons}
\end{center}
\end{figure*}

From Eq.(\ref{k10}), we summarize a counting rule for type-I, type-II and massive Nambu-Goldstone modes:
\begin{align}
&{\rm{type-I}} : {\rm dim} G/H - {\rm rank} \rho \equiv N_\mathrm{I}, \label{k13} \\
&{\rm{type-II}} : \frac{1}{2}~{\rm rank} \rho \equiv N_\mathrm{II}, \label{k14} \\
&{\rm{massive}} : \frac{1}{2} \left( {\rm rank} \bar{g} + {\rm rank} \rho - {\rm dim} G/H \right), \label{k15}
\end{align}
where $\bar{g}(0)$ is defined in Eq. (\ref{k2}) and we used $\mathrm{rank}\bar{g}\equiv \mathrm{rank}\bar{g} (0) =\mathrm{rank}\tilde{\bar{g}} (0)$. Note that
\begin{align}
N_\mathrm{I} + 2 N_\mathrm{II} = N_\mathrm{BG}
\end{align}
with $N_\mathrm{I}, N_\mathrm{II}$ and $N_{\mathrm{BG}}$ the number of type-I and type-II Nambu-Goldstone modes and the broken symmetries is satisfied as reported in Refs.\cite{Hidaka:2012ym,Watanabe:2012hr}. The result is also schematically shown in Fig.\ref{Fig1}. The total degrees of freedom of the system for Nambu-Goldstone modes read
\begin{align}
 N_{\mathrm{NG}}\equiv \frac{1}{2} \left( {\rm dim} G/H +{\rm rank} \bar{g} \right). \label{k16}
\end{align}

We emphasize that the number of Nambu-Goldstone modes is essential for the Higgs mechanism as will be clarified below:  All of the type-I, type-II and massive Nambu-Goldstone modes are ``eaten" by gauge fields. 

\section{Constraints on the effective field theory without Lorentz invariance}
\label{constraints for NG}
The reduction of the number of Nambu-Goldstone bosons is clarified by the Dirac-Bergmann theory of constraints.
The effective Lagrangian without Lorentz invariance may be singular, which means that some of one time-derivative of pion-fields cannot be expressed by the canonical momenta of the pion fields. The canonical momenta $P_{\pi}^\alpha$ are given by
\begin{align}
P_{\pi}^a &\equiv \frac{\partial \mathcal{L}_\pi }{\partial \dot{\tilde{\pi}}^a} \notag \\
&=\frac{1}{2}\rho_{ab}\tilde{\pi}^b +\tilde{g}_{ab}(0)\dot{\tilde{\pi}}^b
=\frac{1}{2}\rho_{ab}\tilde{\pi}^b +\bar{F}^2_a\dot{\tilde{\pi}}^a
\end{align}
In the last equation, we do not take the summation for $a$. If rank$\tilde{g}(0)=\mathrm{dim}G/H$, all of $\dot{\tilde{\pi}}^a$ are replaced by some functions of $P_\pi ^a$ and $\tilde{\pi}^a$ and there are no constraints. This situation corresponds to the case without the $\hat{v}_\beta$-sector in Eq.(\ref{k7}). Type-I Nambu-Goldstone modes and type-II Nambu-Goldstone modes along with the massive ones appear and Type-II Nambu-Goldstone modes without the massive ones do not appear. On the other hand, if $\tilde{g}_{ab}(0)=0$, none of $\dot{\tilde{\pi}}^a$ can be expressed with $P_\pi ^a$ and $\tilde{\pi}^a$ and the number of constraints is dim$G/H$. Thus the number of physical degrees of freedom is $\frac{1}{2}\left(2\mathrm{dim}G/H - \mathrm{dim}G/H \right)=\frac{1}{2}\mathrm{dim}G/H$, corresponding to the case with the only $\hat{v}_\beta$-sector in Eq.(\ref{k7}). The type-II Nambu-Goldstone modes without the massive ones just appear.

Now we consider the general case of rank$\tilde{g}(0)<\mathrm{dim}G/H$ and assume $\bar{F}_{\beta_{(i)}}=0(i=1,2, \beta=1,\dots ,q)$ according to the notation in Sec.\ref{counting}.
 In this case, $2q$ primary constraints appear:
\begin{align}
\phi ^{\beta_{(i)}} =P_{\pi}^{\beta_{(i)}} -\frac{1}{2}\rho_{{\scriptscriptstyle \beta_{(i)}}a}\tilde{\pi}^a .
\end{align}

The total Hamiltonian for the effective Lagrangian without Lorentz invariance is defined by
\begin{align}
\mathcal{H}_\pi &= \dot{\tilde{\pi}}^a P^a_\pi-\mathcal{L}_\pi + \lambda ^{\beta_{(i)}} \phi ^{\beta_{(i)}}\notag \\
&=\frac{1}{2\bar{F}_{\alpha_{(i)}}^2}\left\{P_\pi ^{\alpha_{(i)}} - \frac{1}{2}\rho _{{\scriptscriptstyle \alpha_{(i)}} a}\tilde{\pi}^a \right\}^2+\frac{1}{2\bar{F}_{2p+2q+\gamma}}\bigg\{P_\pi ^{2p+2q+\gamma}\bigg\}^2 \notag \\
&+\frac{1}{2}F_a^2 \partial_r \pi^a \partial_r \pi^a + \lambda ^{\beta_{(i)}} \phi^{\beta_{(i)}} \notag \\
&\equiv \mathcal{H}_\pi ^0  + \lambda ^{\beta_{(i)}} \phi^{\beta_{(i)}}
,
\end{align}
where the summation for $\alpha _{(i)}, \beta_{(i)}, \gamma$ and $a~(i=1,2, \alpha=1, \dots,p, \beta=1,\dots,q, \gamma =1,\dots s$ and $a=1,\dots, \mathrm{dim}G/H)$ are taken due to the notation given by Sec.{\ref{counting}}. In this case, all of the Lagrange multipliers $\lambda ^{\beta_{(i)}}$ are determined due to the Poisson bracket of primary constraints,
\begin{align}
 \left\{\phi ^{\beta_{(i)}},\phi ^{\beta '_{{(j)}}}\right\}_{P} =-{{\rho}}_{{\scriptscriptstyle \beta_{{(i)}}}{\scriptscriptstyle{ \beta'_{(j)}}}},
\end{align}
which means that these constraints are second class. Here we used $\{ \tilde{{\pi}}^a(t,{\bf x}), P_{\pi}^b(t,{\bf y}) \}_P \equiv  \delta^{ab} \delta^3 ({\bf x}-{\bf y})$.
The consistent condition of the primary constraints for the time evolution is given by
\begin{align}
\left\{\phi ^{\beta_{(i)}}, \int d^3 x \mathcal{H}_\pi \right\}_P&\approx
\left\{\phi ^{\beta_{(i)}}, \int d^3 x \mathcal{H}_\pi^0 \right\}_P- {{\rho}}_{{\scriptscriptstyle \beta_{{(i)}}}{\scriptscriptstyle{ \beta'_{(j)}}}}\lambda ^{\beta'_{(j)}} \notag \\
&\approx 0, \label{time_ev}
\end{align}
where ``$\approx$'' is a weak equality, which means an equality under constraint conditions. Particularly, ``$\approx 0$'' is called weakly zero. We can determine all of the multipliers with this equation.

Thus, the physical degrees of freedom are
\begin{align}
&\frac{1}{2}\left\{2 \times \mathrm{dim}G/H - \left(\mathrm{dim}G/H-\mathrm{rank}\tilde{g}(0)\right)\right\} \notag \\
=&\frac{1}{2}\left(\mathrm{dim}G/H +\mathrm{rank}\tilde{g}(0)\right)
=N_{NG},
\end{align}
corresponding to the number of Nambu-Goldstone modes. Note that
 the case of rank ${{\rho}}_{{\scriptscriptstyle \beta_{{(i)}}}{\scriptscriptstyle{ \beta'_{(j)}}}}< 2q$ is more complicated. In this case, some of $\lambda ^{\beta_{(i)}}$ are not determined with Eq.(\ref{time_ev}) and further works are needed.
\section{Higgs mechanism without Lorentz invariance}
\label{constraints for Higgs}
Next we give a general analysis of the Higgs mechanism with the gauged nonrelativistic effective field theory \cite{Leutwyler:1993gf}.
We consider a generic Higgs model composed of the gauged nonrelativistic effective Lagrangian and the kinetic term of gauge fields,
\begin{align}
&\mathcal{L}_{\rm T} = \mathcal{L}_{\pi}^{(0,1)} + \mathcal{L}_{\pi}^{(2,0)} + \mathcal{L}_{\pi}^{(0,2)} + \mathcal{L}_{\rm gauge}, \label{k17} \\
&\mathcal{L}_{\pi}^{(0,1)} = c_a(\pi) \dot{\pi}^a + e_i(\pi) f_0^i,  \label{k18} \\
&\mathcal{L}_{\pi}^{(2,0)} = - \frac{1}{2} g_{ab} (\pi) \partial_{r} \pi^a \partial_{r} \pi^b +h_{ai} (\pi) f_r^i \partial_r \pi^a -\frac{1}{2} k_{ij}(\pi) f_r^i f_r^j, \label{k19} \\
&\mathcal{L}_{\pi}^{(0,2)} = \frac{1}{2} \bar{g}_{ab} (\pi) \dot{\pi}^a \dot{\pi}^b - \bar{h}_{ai} (\pi) f_0^i \dot{\pi}^a + \frac{1}{2} \bar{k}_{ij}(\pi) f_0^i f_0^j, \label{k20} \\
&\mathcal{L}_{\rm gauge} = -\frac{1}{4} F_{\mu \nu}^i F^{i \mu \nu}, \label{k21} 
\end{align}
where $f_0 ^i$ and $f_r^i(i= 1 ... \mathrm{dim} G)$ are gauge fields for time and spatial directions and $F_{\mu \nu}^i$ is the field strength $F_{\mu \nu}^i\equiv \partial _\mu f_\nu ^i - \partial _\nu f_\mu ^i +f^i_{jk}f^j_\mu f^k_\nu$ with the structure constant $ f^i_{jk}$. The coefficients are related through the consistency between the classical equation of the pion fields and the Ward-Takahashi identities \cite{Leutwyler:1993gf,Leutwyler:1993iq}. We note that in addition to the relation in Refs. \cite{Leutwyler:1993gf,Leutwyler:1993iq}, the condition between ${k}_{ij}(\pi)$ and $h^a_i(\pi)$ is also obtained from the consistency,
$h^a_i(\pi) \partial _a {k}_{jk}(\pi)=f^i_{jl} {k}_{lk}(\pi)$.
In particular 
\begin{align}
f^i_{jl} {k}_{lk}(0)=0 \label{add}
\end{align}
for the index $i$ corresponding to the unbroken symmetry $H$.
Using this model, we study the influence of type-I, type-II and massive Nambu-Goldstone bosons on Higgs phenomena. In particular, the physical degrees of freedom are of interest in Higgs phenomena without Lorentz invariance. In the case with Lorentz invariance, the number of Nambu-Goldstone modes coincides with that of pion fields $\pi ^a$ and the number of ``eaten" Nambu-Goldstone modes by gauge fields is that of pion fields. However, in the case without Lorentz invariance, the number of pion fields and the Nambu-Goldstone modes do not always coincide and hence it is not clear whether the number of the ``eaten" Nambu-Goldstone modes coincides with that of the pion fields or Nambu-Goldstone modes. Furthermore, it is not clear at all whether the type-II Nambu-Goldstone modes and massive Nambu-Goldstone modes are absorbed or not.

To clarify the physical degrees of freedom of this system, we adopt unitary gauge, $\pi^a = 0$, where the pion fields are absorbed by gauge fields and analyze the Hamiltonian on the basis of the theory of constrained systems. 
The Lagrangian is reduced to
\begin{align}
&\mathcal{L}_{\rm T} = e_i(0) f_0^i -\frac{1}{2} k_{ij}(0) f_r^i f_r^j + \frac{1}{2} \bar{k}_{ij}(0) f_0^i f_0^j  -\frac{1}{4} F_{\mu \nu}^i F^{i \mu \nu}. \label{k22}
\end{align}
From Eq.(\ref{k22}), we see that the Lagrangian is singular so that we treat it as a constraint system where the method of analysis was formulated by Dirac and Bergmann \cite{Dirac:1950pj}.

First we construct the Hamiltonian of the system.
The canonical momenta are
\begin{align}
\Pi^{i\mu}= F^{i \mu 0}. \label{k23}
\end{align}
$\Pi^{i0} = 0$ is satisfied and thus the primary constraints appear as in Yang-Mills theory:
\begin{align}
\phi^i \equiv \Pi_0^{i}. \label{k24}
\end{align}
The total Hamiltonian is given by
\begin{align}
H_{T} &= \Pi^{ir}\dot{f}_r^i-\mathcal{L}_{\rm T}+ \lambda^i \phi^i \nonumber \\
&= \frac{1}{2} \Pi^{ir} \Pi^{ir} + \Pi^{ir} \left( \partial_r f_0^i + f_{jk}^i f_r^j f_0^k \right) \nonumber \\
&-e_i(0) f_0^i +\frac{1}{2} k_{ij}(0) f_r^i f_r^j -\frac{1}{2} \bar{k}_{ij}(0) f_0^i f_0^j+ \frac{1}{4} F_{rs}^i F_{rs}^i \nonumber \\
&+ \lambda^i \phi^i, \label{k25}
\end{align}
with $\lambda^i$ Lagrange multipliers. 

Next, the consistency of the primary constraints for time evolution yields secondary constraints,
\begin{align}
\chi^{(1)i} &\equiv \Big{\{} \phi^i , \int d^3x H_T \Big{\}}_P \notag \\ 
&\approx \partial_r \Pi^{ir} + f_{jk}^i f_r^j \Pi^{kr} + e_i(0) + \bar{k}_{ij}(0) f_0^j. \label{k26}
\end{align}
Here $\{ \dots \}_P$ is the Poisson bracket given by $\{ f_\mu^{i}(t,{\bf x}), \Pi^{\nu j}(t,{\bf y}) \}_P \equiv \delta_\mu^{\nu} \delta^{ij} \delta^3 ({\bf x}-{\bf y})$ and "$\approx$" is a weak equality, corresponding to an equality under constraint conditions.

With the relation given in Refs. \cite{Leutwyler:1993iq,Leutwyler:1993gf}, $\bar{k}_{ij}(\pi)=\bar{g}_{ab}(\pi) h^a_i(\pi) h^b_j(\pi)$, and Eqs. (\ref{g_gbar}) and (\ref{pion_decay_const}), we can write as
\begin{align}
\bar{k}_{ij}(0)= \left(
\begin{array}{cccccc}
	\bar{F}^2_1&&&&& \\
	&\ddots &&&& \\
	&&\bar{F}^2_{2p+s}&&& \\
	&&& 0&&\\
	&&&&\ddots &\\
	&&&&&0 \\
\end{array}
\right). \label{k27}
\end{align}
Note that ${\rm rank} \bar{k}(0) = {\rm rank} \bar{g}(0)=2p+s$ and the sector with $\bar{F}^2_\delta ~(\delta=1, ... 2p,2p+2q+1,...,2p+2q+s)$ appears due to the sector with two-time derivatives in the matrix $A$ \cite{com2.5}. 

In order to elucidate the consistency conditions for the $\chi^{(1)i}$, we divide the indices to three parts:
\begin{align}
&X=\{ 1,\dots, {\rm rank} \bar{k} \}, \notag \\
&Y_1=\{ {\rm rank} \bar{k}+1,\dots, {\rm dim} G/H \}, \notag \\
&Y_2=\{ {\rm dim} G/H + 1 ,\dots, {\rm dim} G \}, \notag
\end{align} 
with $\mathrm{rank}\bar{k} \equiv \mathrm{rank}\bar{k}(0)$.
Each of $\chi^{(1)i}$ for $i\in X$ and $i\in Y_1$ is related to the terms with and without two-time derivatives in Eq. (\ref{k4}) due to $\bar{k}_{ab}(0) =\tilde{\bar{g}}_{ab}(0)$. $\chi^{(1)i}$ for $i\in Y_2$ corresponds to the part of unbroken symmetry $H$.

In this classification, we can rewrite Eq.(\ref{k26}) as
\begin{align}
\chi^{(1)i}= \left \{ 
\begin{array}{ll}
	\partial_r \Pi^{ir} + f_{jk}^i f_r^j \Pi^{kr} + e_i(0) + \bar{F}^2_i f_0^i,~~(i \in X)  \\
	\partial_r \Pi^{ir} + f_{jk}^i f_r^j \Pi^{kr} + e_i(0).~~~~~~(i \in Y_1 \cup Y_2) \\
\end{array}
\right. \label{k28}
\end{align}
In $i \in X$, the consistency conditions for $\chi^{(1)i}$ are satisfied by determining the Lagrange multiplier, and thus further conditions are not required.
In $i \in Y_1$, on the other hand, the consistency conditions for $\chi^{(1)i}$ give further constraints:
\begin{align}
\chi^{(2)i} \approx f^k_{ij}e_k(0) f_0^j - F^2_i  \partial_r f_r^i-F^2_kf^i_{jk}f_r^jf_r^k.~~~~~(i \in Y_1) \label{k29}
\end{align}
Here $f^k_{ij}e_k(0) = \rho_{ij}$, because $e_i (0)$ corresponds to the expectation value of the charge density \cite{Watanabe:2012hr, Leutwyler:1993gf}, $e_i(0)= \left< 0\right| j^0_i(x) \left|0 \right>$. We note that if the first term is zero for some $i$ in this sector, both one-time and two-time derivative terms for some $\pi^a$ disappear, because $\bar{k}_{ij}(0) = \tilde{\bar{g}}_{ij}(0)= 0$ in $i \in Y_1$. Then the term should be nonzero and further constraints in $i \in Y_1$ are not required.

The consistency conditions for $\chi^{(1)i}$ in $i \in Y_2$ are trivially satisfied by using Eqs.(\ref{pion_decay_const}) and (\ref{add}) as in Yang-Mills theory, which indicates the existence of the remnant gauge symmetry \cite{Castellani:1981us,Sugano:1989rq},
\begin{align}
\Big{\{} \chi^{(1)i} , \int d^3x H_T \Big{\}}_P \approx 0. ~~~~~(i \in Y_2) \label{k31}
\end{align}

From the above discussion, we obtain all of the constraints for the Lagrangian, Eq.(\ref{k22}). Furthermore we classify these into ``first-class " and ``second-class " constraints. In the first-class constraints, all of Poisson brackets with any constraints are weakly zero, while in the second class ones, at least one of Poisson brackets is not zero. By taking the linear combination of all of the constraints and maximizing the number of the first-class ones, we redefine these constraints.

Note that second-class constraints kill one physical degrees of freedom. On the other hand, first-class constraints are related to the gauge symmetry and need additional constraints to fix it. Thus, the first-class ones correspond to kill two physical degrees of freedom \cite{Dirac:1950pj}.

To classify the constraints, we calculate the Poisson brackets:
\begin{align}
&\{ \phi^i , \phi^j \}_P \approx 0~~~~~\left(i,j \in X \cup Y_1 \cup Y_2 \right), \label{k32} \\
&\{ \chi^{(2)i} , \chi^{(2)j} \}_P \approx 0~~~~~\left(i,j \in Y_1 \right), \label{k33} \\
&\{ \chi^{(1)i} , \chi^{(1)j} \}_P \approx - \{ \chi^{(2)i} , \phi^{j} \}_P \approx \left \{
\begin{array}{cc}
	- \rho_{ij}&(i,j \in Y_1) \\
	0&(\mathrm{others}) \\
\end{array}
\right. , \label{k34} \\
&\{ \chi^{(1)i} , \phi^{j} \}_P \approx \left \{
\begin{array}{cc}
	\bar{F}_{i}^2 \delta^{ij}&(i,j \in X) \\
	0&(\mathrm{others}) \\
\end{array}
\right. . \label{k35}
\end{align}
In Eq.(\ref{k34}), we have used the Jacobi identity to derive the first weak equality.
Though $\{ \chi^{(1)i} , \chi^{(2)j} \}_P$ are not equal to $0$ in general, these values do not affect the classification scheme because these terms can be set to $0$ with the redefinition of $\chi ^{(1)i}$ by adding the linear combination of the constraints $\chi ^{(2)i}$ and $\phi^{(1)i}$ for $i \in Y_1$.
From Eqs.(\ref{k32}-\ref{k35}), we conclude each type of constraints:
\begin{align}
&{\rm First~class}: \phi ^i, \chi ^{(1) i}~(i \in Y_2) \\
&{\rm Second~class}: \phi ^i, \chi ^{(1) i}, \chi ^{(2) j}~(i \in X \cup Y_1,j\in Y_1).
\end{align}
Thus we find that the number of the constraints is given as follows:
\begin{align}
&{\rm First~class}: 2\times {\rm dim} H, \\
&{\rm Second~class}: 3\times {\rm dim} G/H - {\rm rank} \bar{k}.
\end{align}

As a result, the physical degrees of freedom in the Higgs system without Lorentz invariance are found to be
\begin{align}
&\frac{1}{2}\left\{2\times4\times{\rm dim} G - 4\times\mathrm{dim}H -\left(3\times\mathrm{dim G/H} - \mathrm{rank} \bar{k} \right) \right\} \nonumber \\
&= \frac{1}{2}\left( \mathrm{dim}G/H+ \mathrm{rank} \bar{k} \right)+ 2\times \mathrm{dim}G \nonumber \\
&=\frac{1}{2}\left( \mathrm{dim}G/H+ \mathrm{rank} \bar{g} \right)+ 2\times \mathrm{dim}G,
\label{k38}
\end{align}
where we used $\mathrm{rank} \bar{k}=\mathrm{rank} \bar{g}$. 
Comparing Eqs. (\ref{k38}) with (\ref{k16}), we conclude the degrees of freedom of the gauge fields in this system coincide with those of the Nambu-Goldstone bosons (the first term) and pure gauge fields (the second term). Thus all of the Nambu-Goldstone modes are absorbed by the gauge fields. This notable fact implies that it is irrelevant for the degrees of freedom in the Higgs system whether the system includes type-I, type-II, or massive Nambu-Goldstone modes, but the existence or non-existence of two-time derivative terms are essential.
Particularly in the system with one-time derivative terms and without two-time derivative terms, the absorbed Nambu-Goldstone modes are just type-II Nambu-Goldstone modes, while in the system with one-time and two-time derivative terms, those are type-II and massive Nambu-Goldstone modes. 

\section{Examples of Higgs models without Lorentz invariance}
\label{examples}
We give two examples of the Higgs systems without Lorentz invariance: a gauged ferromagnet model and SU(2) Higgs-Kibble model with the fixed length of the Higgs field at finite chemical potential $\mu$.

\subsection{gauged ferromagnat system}
\label{GF}
In a ferromagnet system corresponding to the case of $G=$ O(3) and $H=$O(2), the effective Lagrangian does not include the two-time derivative terms and thus one type-II Nambu-Goldstone mode without the massive ones appears \cite{Leutwyler:1993gf, Hofmann:1998pp}.

If the system is gauged, the Lagrangian is given by 
\begin{align}
\mathcal{L} 
   &=-\frac{1}{4}F_{\mu\nu i}F^{\mu\nu i}-\frac{\Sigma}{2}\epsilon^{\alpha \beta}\pi^\alpha \partial_0 \pi^\beta+\Sigma f_0^3+\Sigma \pi ^\alpha f_0^\alpha \nonumber \\
&- \frac{1}{2}F^2\partial_r \pi ^\alpha 
\partial_r \pi ^\alpha -F^2\epsilon ^{\alpha \beta}\partial_r \pi^\alpha f_r^{\beta}-\frac{1}{2}F^2 f_r ^\alpha f_r^{ \alpha}   \label{Gferro_pi_L}
\end{align}
with $\alpha = 1,2, i=1,2,3$. In unitary gauge, $\pi^\alpha =0$, the Lagrangian reduces to 
\begin{align}
\mathcal{L}=-\frac{1}{4}F_{\mu \nu i} F^{\mu \nu i}+\Sigma f_0^3 -\frac{1}{2}F^2f_r ^\alpha f_r ^\alpha . 
\end{align}
We count the absorbed degrees of freedom with the Dirac-Bergmann theory of constraints. The canonical momenta are given by $\Pi^{i\mu}=F^{i\mu 0}$ and the primary constraints are given by $\phi^i = {\Pi_0 ^i}$. The consistency condition for the time evolution of the primary constraints are given by
\begin{align}
\chi^{(1)\alpha}  
&\approx \partial_r \Pi^{\alpha r} + \epsilon_{\alpha \beta} \left(f_r^\beta \Pi^{3r}- f_r^3 \Pi^{\beta r} \right) \notag \\
\chi^{(1)3}  
&\approx \partial_r \Pi^{3r} + \epsilon_{\alpha \beta} f_r^\alpha \Pi^{\beta r} + \Sigma
\end{align}
The consistency condition for $\chi^{(1)3}$ is trivially satisfied, which is related to the remnant gauge symmetry O(2). On the other hand, the consistency conditions for $\chi^{(1)\alpha}$ appear:
\begin{align}
\chi^{(2)\alpha}\approx \epsilon_{\alpha \beta} \Sigma \ f_0^\beta - F^2  \partial_r f_r^\alpha+F^2\epsilon_{\alpha \beta}f_r^\beta f_r^3.
\end{align}
Because of the first term, the consistency conditions for $\chi^{(2)\alpha}$ are satisfied with the determination of the Lagrange multiplier $\lambda ^{\alpha}$. Therefore all constraints are obtained.

Next, we classify these constraints into first-class and second-class constraints. The Possion brackets between these constraints are given as follows:
\begin{align}
&\{ \chi^{(1)\alpha} , \chi^{(1)\beta} \}_P \approx - \{ \chi^{(2)\alpha} , \phi^{\beta} \}_P \approx -\Sigma \epsilon_{\alpha \beta}  , 
\end{align}
and other Possion brackets are zero except for $\{ \chi^{(1)i} , \chi^{(2)\alpha} \}_P \neq 0$. With the redefinition of $\{ \chi^{(1)i}$ by taking the linear combination of $\phi^{\alpha}$,  $\{ \chi^{(1)i} , \chi^{(2)\alpha} \}_P$ reduce to zero.
Thus, the first-class constraints are $\phi^3$ and $\chi^{(1)3}$ and the second-class constraints are $\phi^\alpha , \chi^{(1)\alpha}  
$ and $\chi^{(2)\alpha}.$ The degrees of freedom absorbed by gauge fields are 
\begin{align}
\frac{1}{2}\left(4\times 2\times 3 - 2\times 2 -3\times 2 \right) -2\times 3 = 1,
\end{align}
corresponding to the number of the Nambu-Goldstone mode in this system. This also suggests that $\left(1+2\times3\right)$ degrees of freedom appear in the spectrum at least perturbatively.

\subsection{SU(2) Higgs-Kibble model with the fixed length at finite chemical potential}
If  SU(2) Higgs-Kibble model at finite chemical potential $\mu$ does not include gauge fields, the type-I, type-II and massive Nambu-Goldstone modes appear and the reduction of degrees of freedom does not occur  \cite{Gusynin:2003yu,Hama:2011rt}. By fixing the radial degree of freedom for the scaler field corresponding to the Higgs field, the Lagrangian reduces to the effective Lagrangian in the case of one- and two-time derivatives characterized by $G=$ SU(2) and $H=1$.

Therefore SU(2) Higgs-Kibble model at finite chemical potential $\mu$ without fluctuation of the Higgs field also correspond to one of the model analyzed in Sec.\ref{constraints for Higgs}. In this case, the Lagrangian in unitary gauge may be expressed just by the gauge fields,
\begin{align}
\mathcal{L}=\mu M^2 f_0^3 +\frac{1}{2}M^2 f_\mu ^i f^{\mu i}  -\frac{1}{4}F_{\mu \nu i} F^{\mu \nu i}
\end{align}
with $i=1,2,3.$

Now we consider the constraints of this system.
The primary constraints are given by $\phi^i = {\Pi_0 ^i}$ and the consistency condition for the time evolution of the primary constraints are given by
\begin{align}
\chi^{(1)i}\approx \partial_r \Pi^{i r} + \epsilon_{ijk}f_r^j \Pi^{kr}+\mu M^2 \delta_{i3}+F^2f_0^i.
\end{align}
Note that the last term leads to the determination of the Lagrange multiplier with the consistent condition for the time evolution. This means that all of the constraints,$\phi^i$ and $\chi^{(1)i}$, are second-class ones. Thus the number of absorbed Nambu-Goldstone modes is
\begin{align}
\frac{1}{2}\left(4\times 2\times 3  -2\times 3 \right) -2\times 3 = 3,
\end{align}
corresponding to the total number of type-I, type-II and massive Nambu-Goldstone modes.

\section{Summary and concluding remarks}
\begin{figure*}[h]
\begin{center}
\includegraphics[scale=0.4]{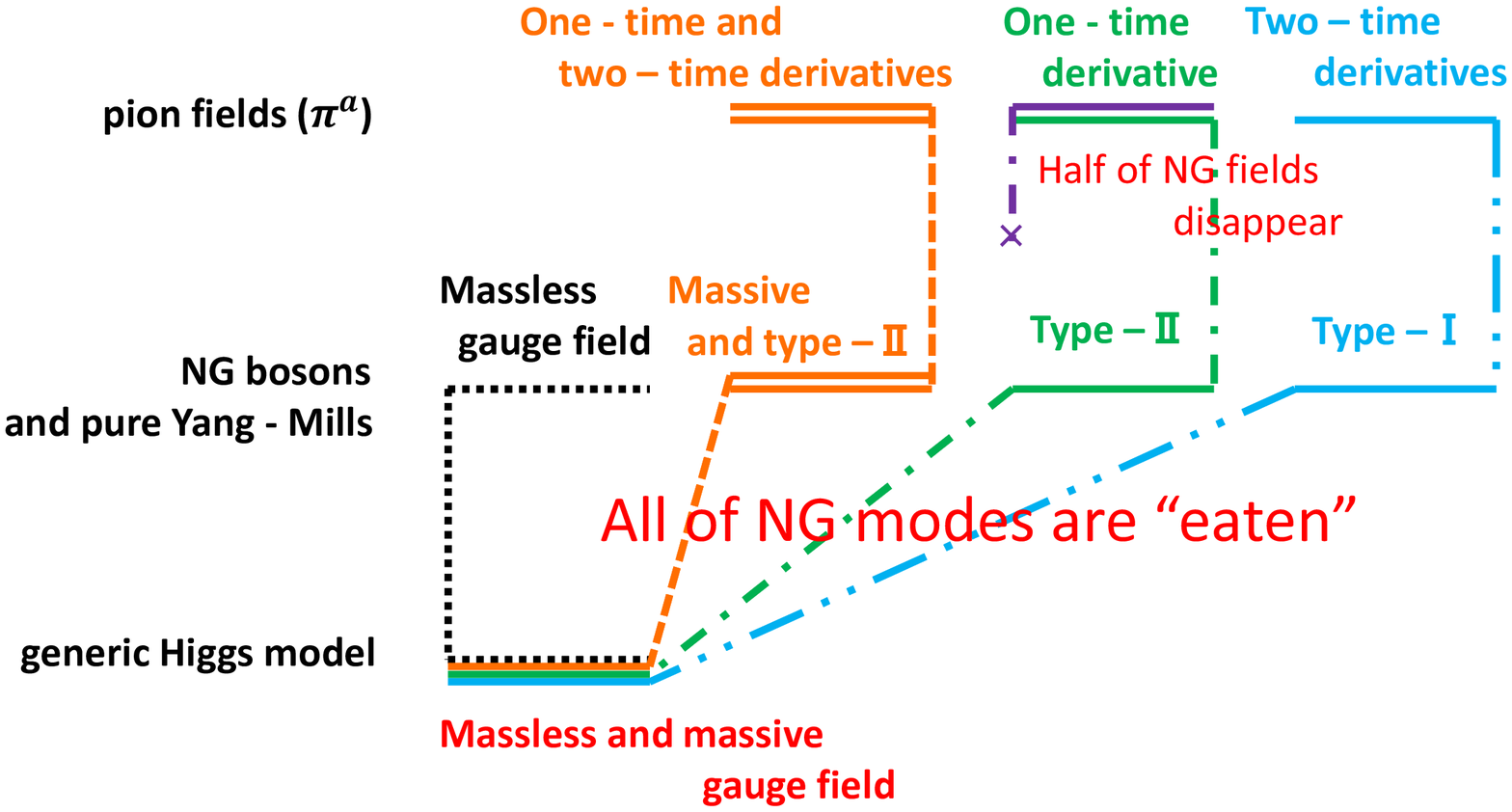}
\caption{\label{Fig2}
The relation between Nambu-Goldstone fields, type-I, type-II and massive Nambu-Goldstone bosons and ``eaten" physical degrees of freedom through the nonrelativistic Higgs mechanism.}
\end{center}
\end{figure*}

In conclusion, we have first elucidated the influence of two-time derivative terms in the effective Lagrangian without Lorentz invariance on the counting rule for Nambu-Goldstone bosons. In this case, type-I, type-II and massive Nambu-Goldstone bosons may appear. In the systems with both one-time and two-time derivative terms, type-II Nambu-Goldstone bosons are accompanied by the massive bosons and the degrees of freedom are not reduced, whereas in the system with one-time and without two-time derivative terms, only type-II bosons appear and the degrees of freedom are reduced. The difference reflects in the Higgs phenomenon in a striking way. We have shown that the number of the eaten physical degrees of freedom by the gauge fields does not coincide with that of the broken symmetries, but that of Nambu-Goldstone bosons, which suggests that these degrees of freedom appear in the spectrum. The schematic illustration of our result is shown in Fig. \ref{Fig2}. 

For future directions, we clarify how gauge fields acquire masses through the nonrelativistic Higgs mechanism by studying the spectrum, which should be helpful to identify the Higgs mechanism without Lorentz invariance in experiment.

\section*{Acknowledgements}
The authors are deeply grateful to Teiji Kunihiro for encouraging and inspiring discussions and his very careful reading of the manuscript and Taichiro Kugo and Yoshimasa Hidaka for the fruitful discussions and helpful comments on the manuscript. S.G. also thanks to Yoshiko Kanada-En'yo and S.K. thanks to Reiji Sugano for valuable discussions. S.~G.~is supported by the Grant-in-Aid for Scientific Research from the JSPS Fellows (No. 24-1458).   

\begin{thebibliography}{99}
\bibitem{com2.5} $\bar{k}$ is redefined by interchanging $(2p+i)$-th row and column with $(2p+2q+i)$-th row and column for $i=1,...,s$.
\end{thebibliography}
\end{document}